\documentclass[
 reprint,
 amsmath,amssymb,
 aps,
]{revtex4-2}

\usepackage[T1]{fontenc}
\usepackage{graphicx}
\usepackage{dcolumn}
\usepackage{bm}
\usepackage{booktabs}
\usepackage{float}
\usepackage{dsfont}
\usepackage{calrsfs}
\usepackage{xcolor}
\usepackage{pict2e}
\usepackage[percent]{overpic}
\usepackage{upgreek}
\usepackage{xfrac}
\usepackage{abraces}
\usepackage{braket}
\usepackage{stmaryrd}
\usepackage{empheq}
\usepackage{multirow}
\usepackage{comment}
\usepackage{listings}

\DeclareMathAlphabet{\pazocal}{OMS}{zplm}{m}{n}
\usepackage[colorlinks=true,linkcolor=blue,citecolor=blue,urlcolor=blue]{hyperref}

\begin{document} 

\preprint{APS/123-QED}

\title{First limits on axion dark matter from a DALI prototype}
\author{Javier De Miguel$^{1,2,3}$}
 \email{jdemiguel@iac.es}


\author{Enrique Joven$^{1}$}%
\author{Elvio Hern\'andez-Su\'arez$^{1}$}%
\author{Juan F. Hern\'andez-Cabrera$^{1}$}%
\author{Haroldo Lorenzo-Hern\'andez$^{1}$}%
\author{Dylan Carroll$^{4}$}%
\author{Roger J. Hoyland$^{1}$}%
\author{Edgar S. Carlin$^{1,2}$}%
\author{Antonios Gardikiotis$^{5,6}$}%
\author{Abaz Kryemadhi$^{7}$}%
\author{J. Daniel Marrero-Falc\'on$^{2}$}%
\author{Marios Maroudas$^{8}$}%
\author{Chiko Otani$^{3}$}%
\author{J. Alberto Rubi\~no-Mart\'in$^{1,2}$}%
\author{Konstantin Zioutas$^{5}$}%

\collaboration{The DALI Collaboration}

\affiliation{$^{1}$Instituto de Astrof\'isica de Canarias, E-38200 La Laguna, Tenerife, Spain}

\affiliation{$^{2}$Departamento de Astrof\'isica, Universidad de La Laguna, E-38206 La Laguna, Tenerife, Spain}

\affiliation{$^{3}$ RIKEN Center for Advanced Photonics,
519-1399 Aramaki-Aoba, Aoba-ku, Sendai, Miyagi 980-0845, Japan}

\affiliation{$^{4}$Independent researcher (unaffiliated); present address: Sagittal Optics, 38002 Santa Cruz de Tenerife, Spain}

\affiliation{$^{5}$Physics Department, University of Patras, GR 26504, Patras-Rio, Greece}

\affiliation{$^{6}$Institute of Quantum Computing and Quantum Technology, National Centre for Scientific Research ``Demokritos'', 153 41 Athens, Greece}

\affiliation{$^{7}$Computing, Math \& Physics, Messiah University, Mechanicsburg, PA 17055, USA}

\affiliation{$^{8}$Institute for Experimental Physics, University of Hamburg, 22761 Hamburg, Germany}

\date{\today}

\begin{abstract}
We report a pilot dark matter search with a cryogenic, magnetized, scaled-down DALI prototype. An analysis of 36 hours of data reveals no statistically significant excess attributable to axionlike particles. We therefore set new exclusion limits in the 6.88691--6.91792 GHz band, reaching an axion–photon coupling sensitivity of $g_{a\gamma\gamma}\lesssim 9.7\times10^{-12}\,\mathrm{GeV}^{-1}$ at $28.54 \, \upmu\mathrm{eV}$. These results consolidate the DALI approach and motivate a next-stage haloscope to explore a broader mass range with upgraded instrumentation.

\end{abstract}

\maketitle



\section{Introduction}\label{sec:intro}
Axions were proposed to address the quantum chromodynamics (QCD) problem of charge-conjugation and parity symmetry in the strong interaction \cite{PhysRevLett.38.1440}. Both axions \cite{PhysRevLett.40.223,PhysRevLett.40.279} and axionlike particles beyond the QCD framework—hereafter collectively referred to as axions—have long been hypothesized to constitute dark matter (DM) \cite{1933AcHPh...6..110Z,ABBOTT1983133,DINE1983137,PRESKILL1983127}, an elusive substance whose existence is supported by indirect evidence \cite{1970ApJ...159..379R}. Exploiting the weak coupling of axions to standard particles, various experimental methods have been developed to search for DM—see Refs. \cite{Arza:2026rsl,Andrieu:2025xpv} for a review.

The axion—photon interaction is described by the effective Lagrangian density
\begin{equation}
\pazocal{L} \supset g_{a\gamma\gamma} a \, \mathbf{E} \cdot \mathbf{B} \,, 
\end{equation}
where $g_{a\gamma\gamma}$ denotes the axion--photon coupling, $a$ is the axion field, $\mathbf{E}$ represents the electric field of the photon, and $\mathbf{B}$ is an external static magnetic field that supplies the virtual photon required for Primakoff conversion \cite{Primakoff:1951iae}. A promising experimental approach to probing this interaction is the haloscope, first proposed by Sikivie \cite{1983PhRvL..51.1415S}, which in its most common implementation employs a magnetized resonant cavity to enhance the weak microwave signal induced by virialized axions in the Galactic halo. Despite several decades of development, the cavity haloscope faces increasing challenges in maintaining high sensitivity at progressively higher frequencies, as wave coherence requirements reduce the effective detection volume, and thus, the achievable sensitivity. In this manuscript we report the first results from a demonstrator of the Dark-photons \& Axion-Like particles Interferometer (DALI), a new-generation DM haloscope designed to complement axion searches at high frequencies.
\begin{figure}[H]
    \centering    \includegraphics[width=.48\textwidth,trim=2.5cm 0 2.5cm 0,clip]{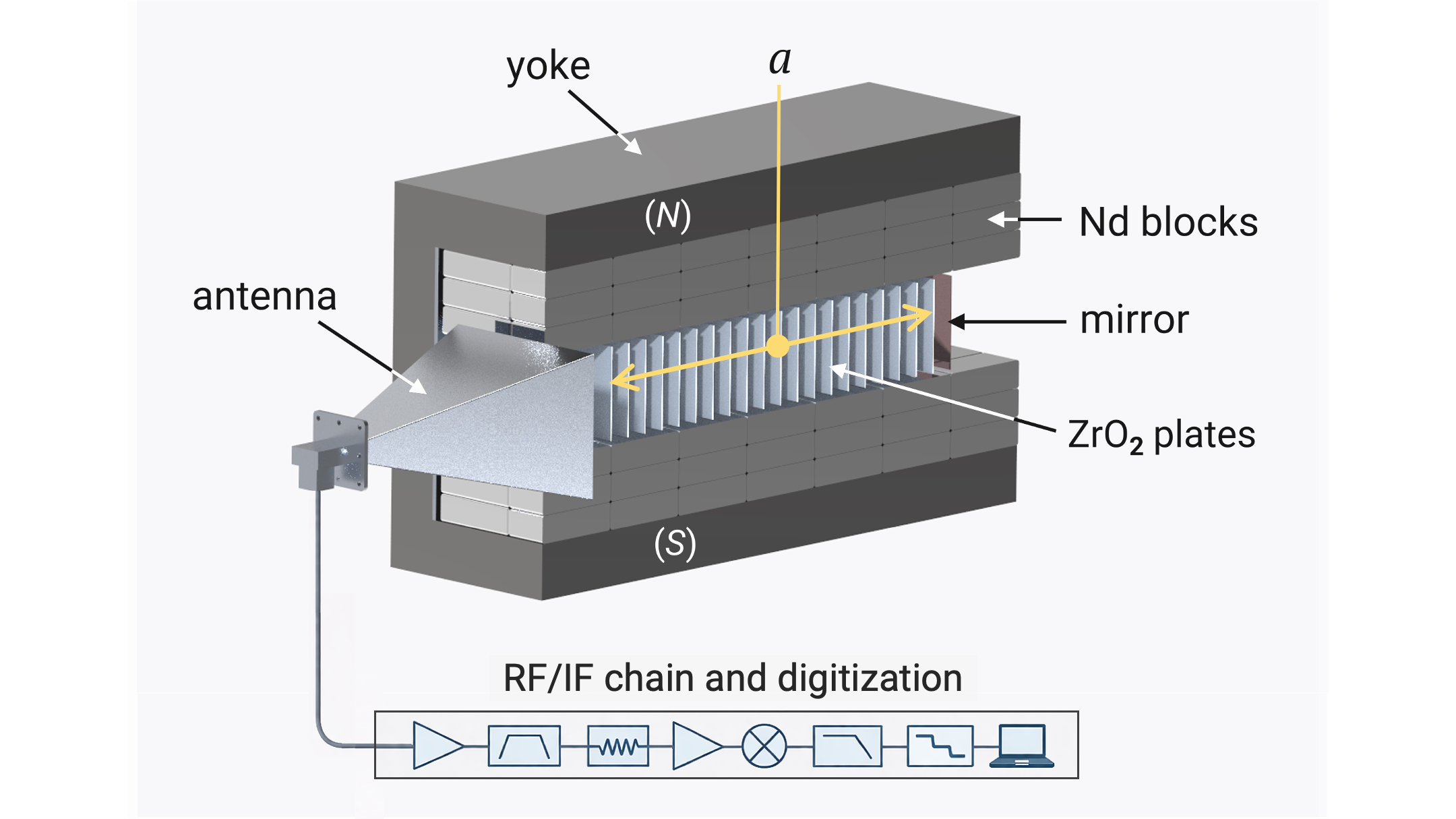}
    \caption{Schematic view of the DALI(PoP) prototype. A tunnel yoke holds an array of Nd blocks that generates a dipole-like magnetic field. Ambient axions are converted into microwave photons, which are collected by the antenna. The resonator, composed of twenty ZrO$_2$ layers and enclosed by a copper mirror, enhances the weak axion-induced signal to an observable level. The detected signal is subsequently processed by the readout chain shown below (from left to right: amplifier, band-pass filter, attenuator, amplifier, downconverter, low-pass filter, analog-to-digital converter, and data storage). The haloscope is housed in a Faraday cage to suppress external spurious signals (not shown). The magnet is segmented along its longitudinal axis.}
    \label{fig_0}
\end{figure}

DALI~\cite{DeMiguel2021,DeMiguel:2023nmz,Cabrera2023qkt,2024JInst19P1022H,PhysRevD.110.072013,DeMiguel:2024cwb} features a novel experimental setup for detecting wavelike DM, known as a \textit{magnetized phased array} (MPA). A resonator placed in a static magnetic field enhances the faint signal expected from cosmic axions, for which a tunable Fabry--P\'erot (FP) interferometer is employed~\cite{1899ApJ.....9...87P}. In the MPA haloscope, the phased array forms a coherent electromagnetic pattern that couples to the resonant field of the FP tuner in the radiative near field, thereby collecting the signal over the open-cavity aperture. Interestingly, the resonance frequency is decoupled from the transverse area of the ceramic plates, allowing the resonator cross section to be scaled independently of frequency. As a result, this architecture offers a complementary path toward resonant axion searches at higher frequencies, where maintaining large effective volumes and high sensitivity becomes increasingly challenging for conventional cavity haloscopes. Notably, DALI relies on mature, readily available components, enabling rapid, cost-effective development; the use of a conventional solenoid-type superconducting magnet, widely deployed in the medical industry, further reduces technical risk and facilitates procurement.

The DALI program comprises different stages. As part of its scientific plan, in this work we report a proof-of-principle (PoP) test at a lower scale, designed to retain a part of DALI's physics potential while relieving hardware heaviness. The rest of this paper is structured as follows. The section \textit{Experimental setup} describes the DALI(PoP) prototype employed to search for axion DM, while \textit{Data analysis} details the observations and their statistical treatment. A new limit on the DM axion parameter space is presented in \textit{Results}.
\section{Experimental setup} 
A schematic view of the DALI(PoP) prototype hosted at the IAC facilities at latitude 28$^\circ$28'29" N and longitude 16$^\circ$18'37" W is shown in Fig. \ref{fig_0}. The apparatus can operate in two frequency ranges, 6--8~GHz and $\sim$30--40~GHz. The PoP prototype scales down DALI by approximately halving the number of layers of the FP tuner and reducing the plate size to $\sim$1:10, which diminishes the physical area by $\sim$100. Thus, while the full-scale MPA is designed to cover larger cross-sectional areas by means of a phased array in which multiple antennas combine coherently, the reduced surface of the prototype can be covered with a single antenna while preserving the same DALI operating principle.
\begin{widetext}
\begin{equation}
\begin{split}
g_{a\gamma\gamma} \gtrsim\;& 2 \times10^{-11}\,\mathrm{GeV^{-1}}
\left(\frac{\mathrm{SNR}}{3}\right)^{1/2}
\left(\frac{10^{3}}{Q_L}\right)^{1/2}
\left(\frac{1+\beta}{\beta}\right)^{1/2}
\left(\frac{100\,\mathrm{cm}^2}{A_{\eta}}\right)^{1/2}\\
&\times
\left(\frac{m_a}{50\,\upmu\mathrm{eV}}\right)^{5/4}
\left(\frac{1\,\mathrm{day}}{t}\right)^{1/4}
\left(\frac{T_\mathrm{sys}}{50\,\mathrm{K}}\right)^{1/2}
\frac{0.6\,\mathrm{T}}{B}
\left(\frac{0.45\,\mathrm{GeV\,cm^{-3}}}{\rho_a}\right)^{1/2}.
\end{split}
\label{Eq.1}
\end{equation}
\end{widetext}
A detailed derivation for the DALI formalism is given in Refs.~\cite{DeMiguel:2023nmz, DeMiguel:2024cwb}. The on-resonance axion-induced signal power in a DALI magnetized interferometer scales as $P_a \propto g_{a\gamma\gamma}^2 B^2 A_{\eta} \, Q_L \, \beta/(1+\beta)$, where $A_{\eta}$ represents the electromagnetically effective
cross-sectional area overlapping the axion-induced source distribution, $Q_L$ denotes the loaded quality factor, $B$ is the external static magnetic field, and $\beta$ accounts for receiver coupling. Combining this scaling with the Dicke radiometer relation \cite{1946RScI...17..268D}, ${\rm SNR}\propto P_a T_{\rm sys}^{-1}(t/\Delta \nu_a)^{1/2}$, with $T_{\rm sys}$ the system noise temperature and $\Delta \nu_a$ the signal linewidth, yields the sensitivity law in Eq. \ref{Eq.1}; where $m_a$ is the axion mass, $t$ is the integration time, and $\rho_{a}$ denotes the local density of DM, set at $0.45^{+0.03}_{-0.09}$ GeV cm$^{-3}$ \cite{Staudt:2024tdq} under the assumption that the axion is its fundamental component. 
\begin{figure}[b]
    \centering    \includegraphics[width=.48\textwidth]{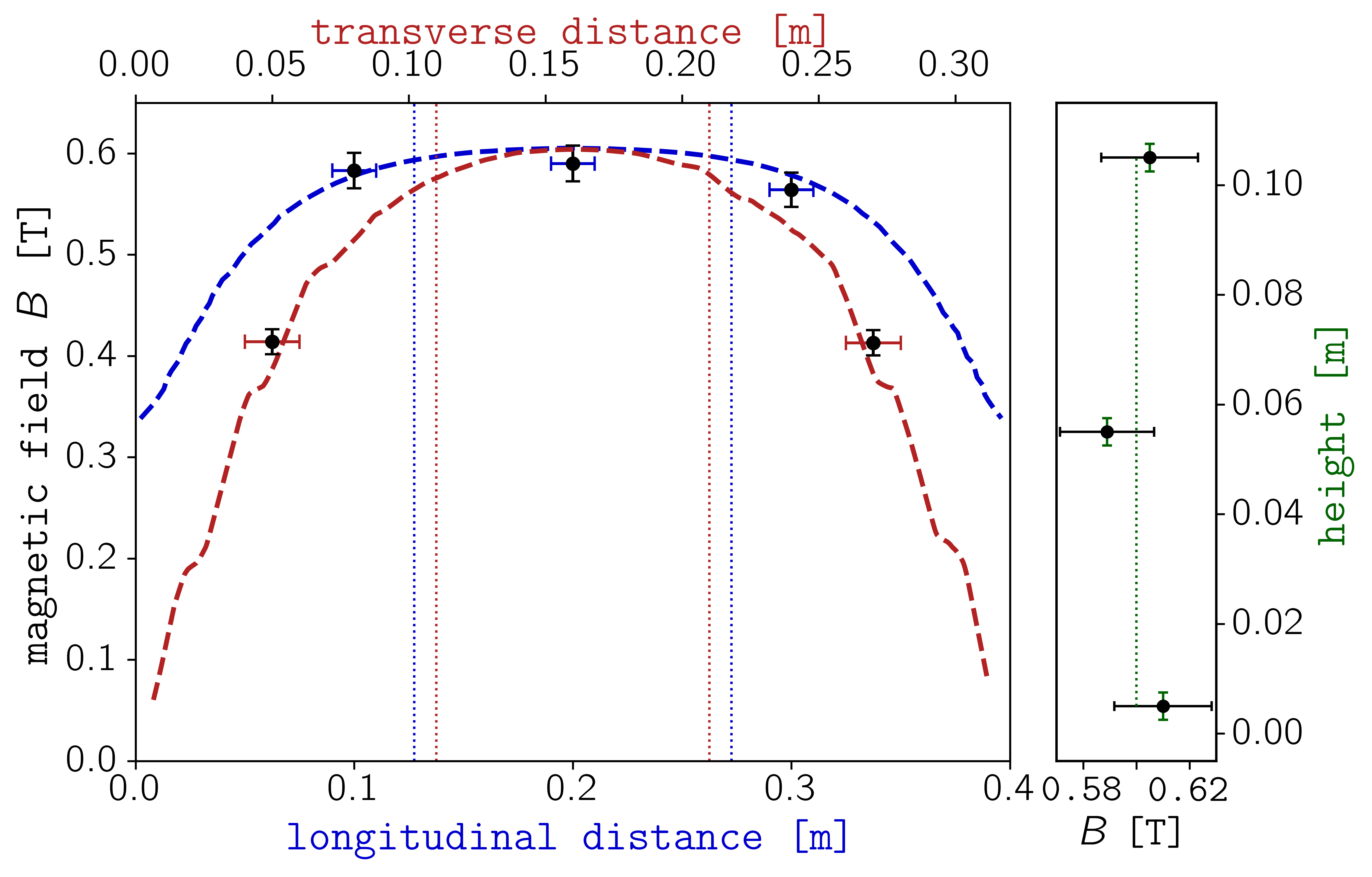}
    \caption{Field simulation \textit{vs.} measurements. The magnet bore is approximately $30$ (transverse)$\times 11$ (height)$\times 40$ (longitudinal)~cm. Two one-dimensional profiles are extracted from the three-dimensional magnetic-field simulations: a center--longitudinal cut at half height (blue dashed) and a center--transverse cut (red dashed). The field generated by the magnets installed inside the steel yoke was simulated using \texttt{Elmer FEM}\,\cite{elmer}, employing \texttt{Gmsh}\,\cite{gmsh} to generate the three-dimensional meshes. The dotted lines indicate the resonator size (10$\times$10$\times$14.5 cm$^3$) in its centered configuration, shown for reference. Control Hall-probe measurements are shown with 3\% vertical error bars and horizontal probe-positioning bars; the right inset displays three measurements along the bore height at the longitudinal center. The simulated field distribution is in good agreement with the measurements.}
    \label{fig_5}
\end{figure}

\begin{figure}[b]
    \centering    \includegraphics[width=.39\textwidth]{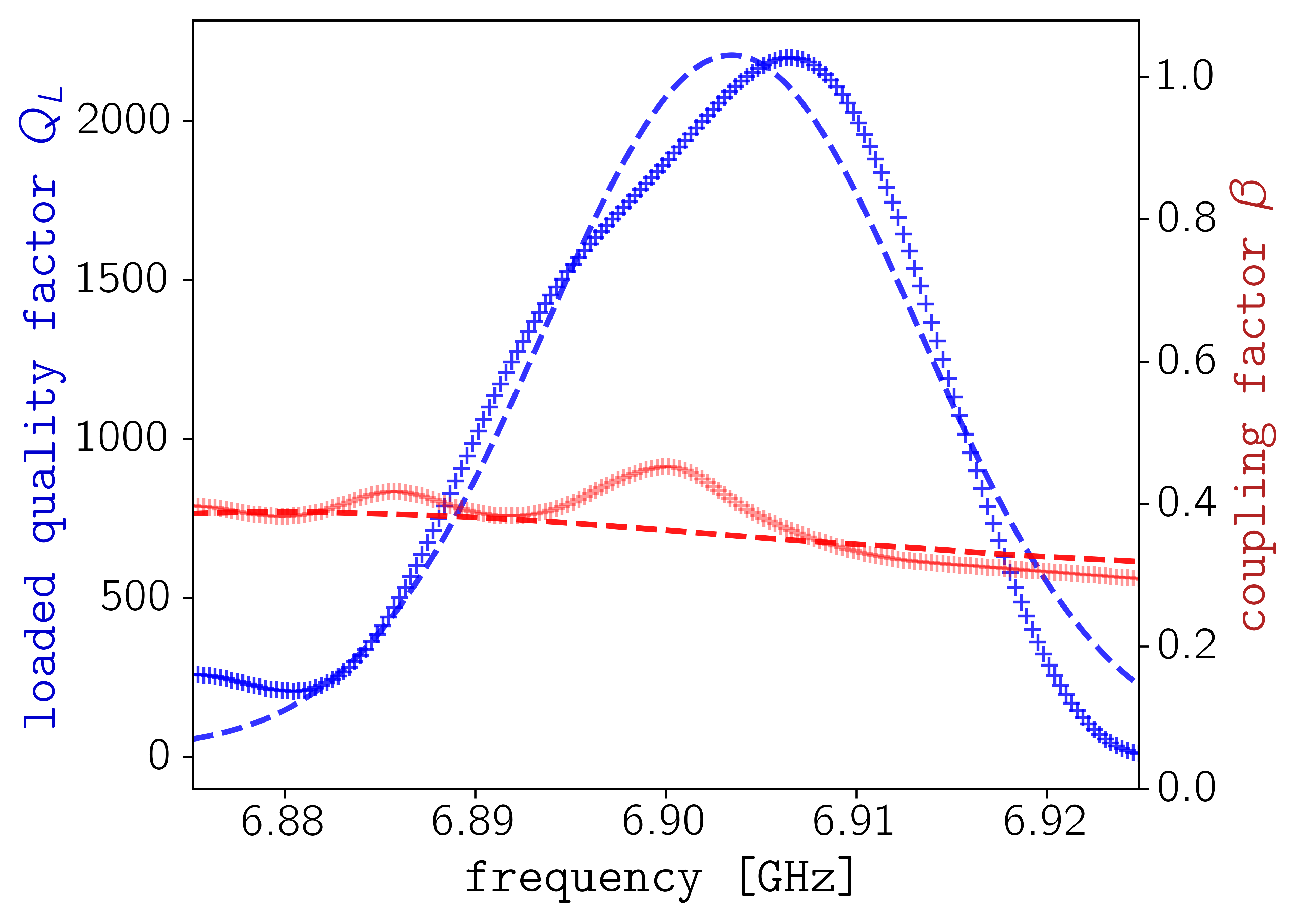}
    \caption{Quality factor and coupling factor. Measured quality factor data are shown in blue. The full width at half maximum for a fit, represented by a blue dashed line, is $\sim$20 MHz. In the present prototype configuration the receiver is undercoupled. The observed optical power coupling factor $\beta$ is depicted in red, while broadband data ($\sim$5--8 GHz) are smoothed with a SG baseline using a window size $W=201$ and polynomial order $d=1$, shown by a red dashed line as a reference. Throughout this work, $Q_L$ and $\beta$ are taken from the measured data, not from the fitted or smoothed curves.}
    \label{fig_4}
\end{figure}

In the DALI(PoP) haloscope, the superconducting solenoid of the full-scale DALI design is replaced by a lighter permanent-magnet assembly consisting of two sets of 36 grade-N44 neodymium (Nd) alloy blocks, each measuring 100$\times$100$\times$30 mm, for a total of 24 three-magnet stacks, 12 at each pole, enclosed in a tunnel-yoke made of F-114 steel sheets with length, width, and height of about 1/2 m and a bore of approximately 30$\times$11 cm$^2$. The magnetic field exhibits $\gtrsim$$95\%$ homogeneity over the resonator volume, as measured with a HIRST-GM08 Hall-effect sensor, with a field strength of $0.59\pm0.01$ T—see Fig. \ref{fig_5}. We employ a FP resonator composed of 20 layers of yttria-stabilized zirconia (ZrO$_2$), each of size 100$\times$100 mm with $\pm$0.5 mm tolerance, thickness 1$\pm$0.03 mm, and surface roughness below 0.1 $\upmu$m; the material has permittivity $\varepsilon_r\sim30$ and dielectric loss tangent $\mathord{\mathrm{tan}}\,\delta\sim 10^{-4}$ at centimeter wavelengths. In this pilot run, the plate spacing was fixed for a single frequency step at $6.21 \pm 0.05$ mm. The loaded quality factor was measured following the laboratory procedure described in Ref.~\cite{PhysRevD.110.072013} with a Rohde \& Schwarz ZNB 20 vector network analyzer (VNA). For the present resonator stack, with an inter-plate spacing of $\sim\lambda/8$ (where $\lambda$ is the wavelength), the response peaks at $Q_L\approx2200$ near 6.907 GHz and follows a pseudo-Lorentzian profile, as shown in Fig.~\ref{fig_4}. 
Microwave absorbers are used to buffer against spillover and suppress reflections—EA-PF3000-24 and 3650-40-ML. A WR-137 standard horn antenna receives the signal, which is sent to a high electron mobility transistor radiometer composed of a cryogenic LNF-LNC4-8C low-noise amplifier, with $\sim$40 dB gain and a noise temperature of a few kelvin, and a room-temperature backend with a second LNF-LNC4-8C, a 12 dB attenuator (Mini-Circuits\textsuperscript{\textregistered} model 15542) to avoid compression at the second gain stage, and $\sim$6--8 GHz band-pass filtering. Powered by two closed-cycle $^{4}$He cryocoolers, the system temperature was determined with the standard Y-factor method by heating a microwave absorber with a resistor (Ohmite HS100 330R J) over $\Delta T \sim 60$ K, yielding $T_{\rm sys}\simeq 59$ K for the present implementation, including contributions from the antenna temperature (49.3 K, cooling-limited under the $>$$300$ kg magnet load) and the receiver noise (9.7 K).

The electromagnetically effective cross section entering Eq.~\ref{Eq.1} was obtained from the simulated transverse modal overlap with the ceramic plate area, giving $A_{\eta}\approx A_{\rm phys} \times 0.61$. The antenna coupling was extracted from free-space VNA measurements; the resulting $\beta < 1$ (see Fig. \ref{fig_4}) reflects an undercoupled antenna--resonator arrangement, leaving room for improved matching in future realizations. The instrumental parameters and uncertainties for this run are listed in Table~\ref{table9}.

\begin{table}[h]
\caption{Uncertainties are weighted according to their order in Eq.~\ref{Eq.1} to estimate the propagated cumulative uncertainty in this run.}
\label{table9}
\begin{tabular}{ccc}
\\ \toprule
Parameter & Value & Uncertainty \\\midrule
Loaded quality factor ($Q_L$) [peak] & 2200 & 8\%\\
Coupling ($\beta$) [at resonance] & 0.36 & 1\%\\
Effective area ($A_{\eta}$) & 61 cm$^2$ & 6\%\\
System temperature ($T_{\mathrm{sys}})$ & 59.0 K & 2\%\\
Magnetic field ($B$) & 0.59 T & 3\%\\\hline
Propagation to $g_{a\gamma\gamma}$ sens. & & 6\%\\
\bottomrule
\end{tabular}
\end{table}

Components near the magnet are made of non-magnetic materials or placed where the fringe field is negligible. The cryostat is enclosed within a Faraday cage from Holland Shielding Systems B.V., equipped with a connector plate fitted with radio-frequency (RF) filters, waveguides, and ventilation honeycombs, to attenuate spurious microwave signals from outside by more than 60 dB. The data acquisition system is built outside the shielding tent and consists of a zero intermediate frequency (ZIF) downconverter model thinkRF\textsuperscript{TM} R5550-408, a two-channel digitizer, and a local server-based storage system. The downconverter shifts the radiometer output from RF to baseband, creating in-phase and quadrature (IQ) outputs for instantaneous acquisition of the full band, with a minor loss of data close to zero hertz. The IQ outputs pass through SLP-36+ low-pass filters with a nominal $-3$ dB cutoff frequency of approximately 40 MHz before simultaneous digitization. A mask balances the I and Q signals in amplitude and phase before channel combination, calibrated with a swept VNA measurement at the input of the ZIF downconverter. Direct acquisition of the IQ baseband signal is enabled by a dual-channel, 14-bit analog-to-digital converter (GaGe CSE8329). The digitized raw signals are continuously streamed to the local storage system via a peripheral component interconnect express link, where the 14-bit analog-to-digital converter output is recorded in 16-bit integer format and can store up to 18 TB of time-domain samples for subsequent post-processing.

\section{Data analysis} 
Data were acquired for 36 hours effective time between February 17 and March 5, 2026, using the equipment described in the previous section. A total of N=12960 spectra were recorded at 125 MS/s, centered at 6.90 GHz. Each stored spectrum corresponds to the average of 9540 consecutive dual-channel fast Fourier transform segments of length 2$^{17}$ samples, yielding an effective integration time of 10 s per file and a frequency resolution of $\Delta\nu_b \simeq 954$ Hz. In total, this corresponds to approximately 64.8 TB of raw time-domain data, compressed to 44.2 GB in the stored frequency-domain dataset.

In Ref.~\cite{Cabrera2023qkt}, we adapted the analysis method of Ref.~\cite{PhysRevD.96.123008} to our MPA haloscope and used the resulting pipeline to perform end-to-end Monte Carlo simulations, in which a synthetic axion signal was injected into a noisy background and its detection rate was used to fine-tune the prefactor in Eq.~\ref{Eq.1}. In this section, we follow the methodology and nomenclature introduced there. First, time-to-frequency-domain conversions were calculated using the \texttt{Fastest Fourier Transform in the West} library \cite{1386650}, and
the spectra were cropped to the region with $Q_L\gtrsim600$, chosen to retain the
sensitive resonant band. The cropped spectra were averaged in the frequency
domain and the resulting spectrum was smoothed using a Savitzky--Golay (SG) baseline with $W=251$ and
polynomial degree $d=4$. At this stage, spurious RF interferences in each
normalized spectrum were identified iteratively as bins exceeding
$5.5\,\sigma$ with respect to that spectrum's mean and standard deviation.
The compromised bins were replaced with random values drawn from the
corresponding Gaussian distribution until no further outliers were found
\cite{doi:10.1126/sciadv.abq3765}; their frequencies were subsequently
masked in the analysis. The residual broad spectral structure in each spectrum
was then removed using a SG filter with $W=251$ and $d=4$. Processed spectra were rescaled by the effective
quality factor, $Q_{\rm eff}=Q_L\beta/(1+\beta)$, at each bin. The rescaled
spectra were then combined and rebinned with parameters $K_r=3$, $K_g=5$,
and $z=0.65$, such that the expected axion line shape spanned approximately
$K_g$ bins. The resulting grand spectrum (GS), with frequency-bin width
$\Delta\nu_{\rm GS}\simeq2.86$ kHz, was obtained from a weighted sum that
accounts for the expected axion line shape in the laboratory frame,
\begin{multline}
f(v) = \frac{2}{\sqrt{\pi}} \left( \sqrt{\frac{3}{2}} \frac{1}{r\, v_a \langle \beta_{\rm rms}^2 \rangle} \right)
\sinh\left( \frac{3r}{\sqrt{2}} \frac{(v - v_a)}{v_a \langle \beta_{\rm rms}^2 \rangle} \right) \\
\times \exp\left[ -\frac{3}{2} \left( \frac{(v - v_a)}{v_a \langle \beta_{\rm rms}^2 \rangle} \right)^2
- \frac{3r^2}{2} \right] \,,
\end{multline}
where \( v \) is the axion velocity in the lab frame, \( v_a \) is the mean axion velocity in the Galactic rest frame, \( r \equiv v_{\mathrm{lab}} / v_\mathrm{rms} \) encodes the lab motion relative to the halo, and \( \langle \beta_{\rm rms}^2 \rangle = 3/2 \, v_\mathrm{rms}^2 / c^2 \) is the mean squared velocity in units of the speed of light \( c \), with \( v_\mathrm{rms} \simeq 270\,\mathrm{km/s} \). The total lab-frame velocity is \( v_{\mathrm{lab}} = v_{\odot} + v_{\oplus} + v_R \), where \( v_{\odot} \simeq 230\,\mathrm{km/s} \) is the Sun’s velocity relative to the Galactic frame, \( v_{\oplus} \simeq 29.8\,\mathrm{km/s} \) is Earth’s orbital speed, and \( v_R = \omega_{\oplus} R_{\oplus} \cos\phi_\mathrm{det} \) accounts for Earth’s rotation at detector latitude \( \phi_\mathrm{det} \), with \( \omega_{\oplus} = 2\pi / \mathrm{day} \) and \( R_{\oplus} \simeq 6371\,\mathrm{km} \) the Earth’s equatorial radius \cite{PhysRevD.96.123008}. The filter-induced correlations among bins in the normalized GS were corrected at first order by dividing by its standard deviation. Candidates were then flagged as the bins exceeding a significance threshold ($\alpha$) for an arbitrary confidence level (CL) assuming a normal distribution as $\alpha=\mathrm{SNR} - \mathrm{CL}$. 

\begin{figure}[t]
    \centering    \includegraphics[width=.48\textwidth]{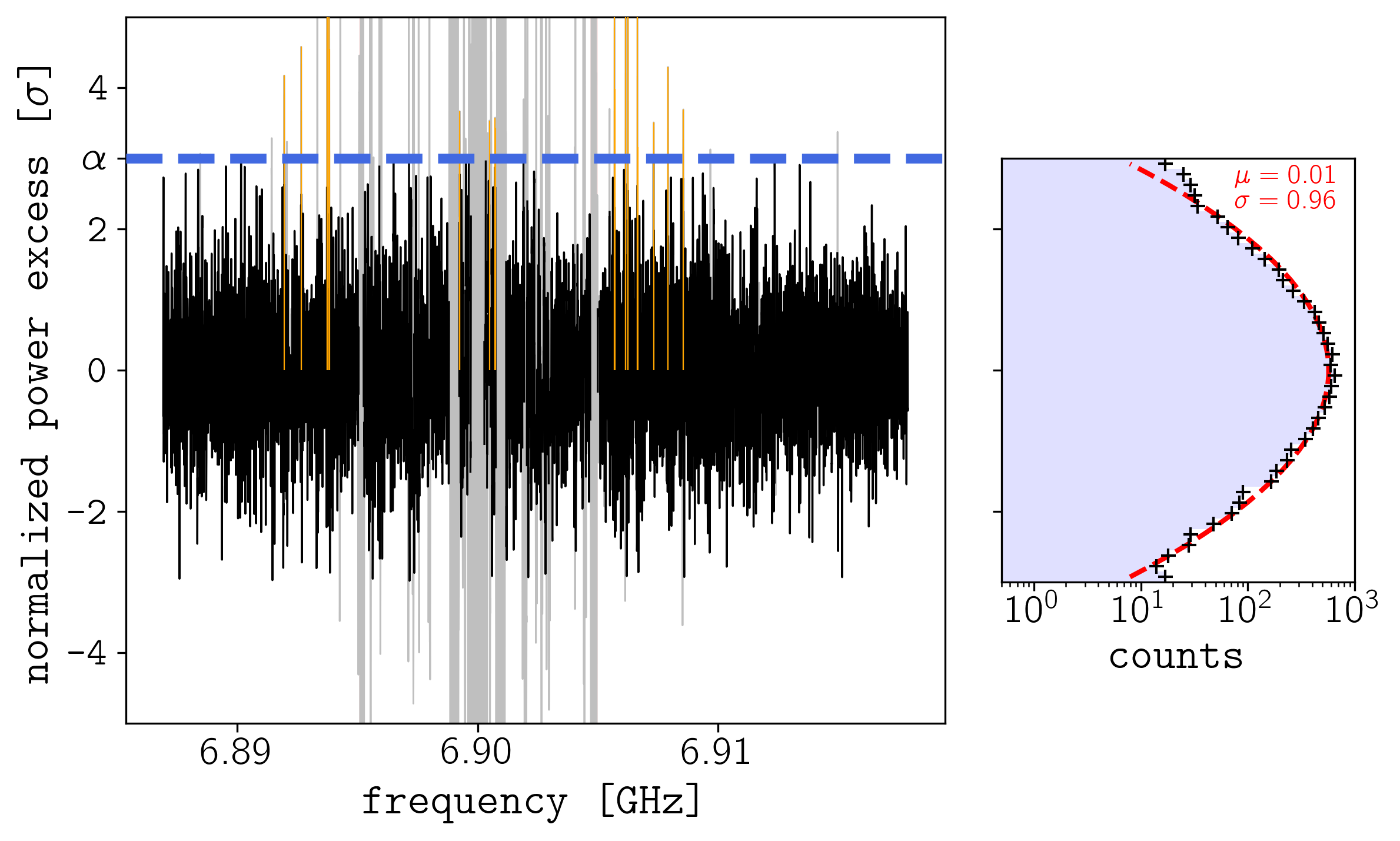}
    \caption{Normalized, correlation-corrected grand spectrum (left). The black trace shows the retained bins, while gray traces denote masked bins; orange traces mark positive clusters flagged for rescan (see Table \ref{TableII}). The blue dashed line indicates the detection threshold, $\alpha=3$. Right: histogram of the retained normalized power excesses, compared with a unit normal reference (red dashed). The grand spectrum is normalized by a robust (median) estimate of the noise scale over the instrumentally unmasked band, so that $\alpha=3$ corresponds to $3\sigma$; the measured $\hat\sigma=0.96$ of the retained bins reflects the symmetric vetoing of positive and negative excursions.}
\label{fig_1}
\end{figure}

\begin{table}[b]
  \centering
\caption{Positive-excess clusters retained for follow-up rescans. Columns list the frequency range and cluster width;
CL$_{\rm loc}$ is the local exclusion confidence level for each cluster.}
\label{TableII}
\begin{tabular}{ccc}
\hline\hline
Frequency [GHz] & $N_{\rm bins}$ & CL$_{\rm loc}$ [\%] \\
  \hline
  6.891944--6.891950 & 3 & 54.2 \\
  6.892648--6.892656 & 4 & 38.1 \\
  6.893729--6.893741 & 5 & 17.3 \\
  6.893818--6.893829 & 5 & 5.3 \\
  6.899245--6.899260 & 6 & 72.9 \\
  6.900493--6.900499 & 3 & 77.2 \\
  6.900730--6.900739 & 4 & 75.9 \\
  6.905700--6.905711 & 5 & 16.5 \\
  6.906163--6.906175 & 5 & 7.4 \\
  6.906252--6.906263 & 5 & 5.2 \\
  6.906655--6.906667 & 5 & 3.6 \\
  6.907334--6.907339 & 3 & 77.9 \\
  6.907929--6.907937 & 4 & 49.4 \\
  6.908564--6.908572 & 4 & 72.1 \\
  \hline\hline
\end{tabular}
\end{table}

Narrow spectral features not removed by the upstream screening can become
significant only at the GS stage, where persistent lines accumulate through
stacking, rebinning, and line-shape weighting; their extent is then assessed
with an adaptive veto procedure. The normalized GS was corrected for its
measured correlation factor, determined from a line-free window. Fixed instrumental masks and bins associated with narrow IF/RF interferences were excluded before candidate selection. Positive excesses
above the threshold $\alpha=3$ were grouped into contiguous clusters, and those
with widths compatible with the expected axion linewidth
($3\leq N_{\rm bins}\leq7$) were provisionally retained; wider clusters were vetoed as broad noise or systematics. Candidate and negative-excursion bins
were excluded using adaptive veto windows.
The end-to-end response was calibrated by injecting the laboratory-frame axion
template into the recorded spectra before the baseline was estimated, so that
the baseline response to the signal itself is included. Injections at five frequencies spanning the band, and at three couplings gave, an end-to-end pipeline detection efficiency of
$\sim 0.6$ ($\eta_{\rm e2e} \simeq 0.6 = \eta_{\rm match}\times\eta_{\rm SG} \simeq 0.76 \times 0.79$) with no measurable dispersion in either variable.
Surviving clusters were subjected to two independent tests. The first is a
line-shape test on the coadded, unrebinned spectrum. We fit a
Gaussian of free width, propagated through the same baseline-removal operator
as the data, and compared it with the width recovered by the axion template propagated identically. The test was applied only where the local excess exceeded five times the median-absolute-deviation noise level of the coadded spectrum, and was otherwise declared
inconclusive rather than used to veto. Secondly, excesses incompatible with the CAST limit at 95\% confidence were flagged and excluded from further consideration. Persistent narrow lines near 6.9 GHz were also observed with the RF input from the cryogenic front-end disconnected and the room-temperature chain terminated with a matched load, and in control runs with the RF center frequency shifted by $+50$~MHz, supporting their instrumental classification. Fourteen of the forty-six width-selected clusters satisfy the criteria and have a local confidence level CL$_{\rm loc}>1$\%, as listed in Table~\ref{TableII}. Negative excursions were vetoed symmetrically with the positive side because baseline removal turns any narrow feature into a residual with
negative sidelobes, so a strong spurious line imprints companion excursions. They
sit more than twice as close to the positive excesses as random placement would
allow, confirming that they trace the same instrumental contamination as the excess
that generates them. The masked regions appear as gray traces and the retained rescan candidates as orange traces in Fig.~\ref{fig_1}.

\section{Results} 
\begin{figure}[H]
    \centering    \includegraphics[width=0.48\textwidth]{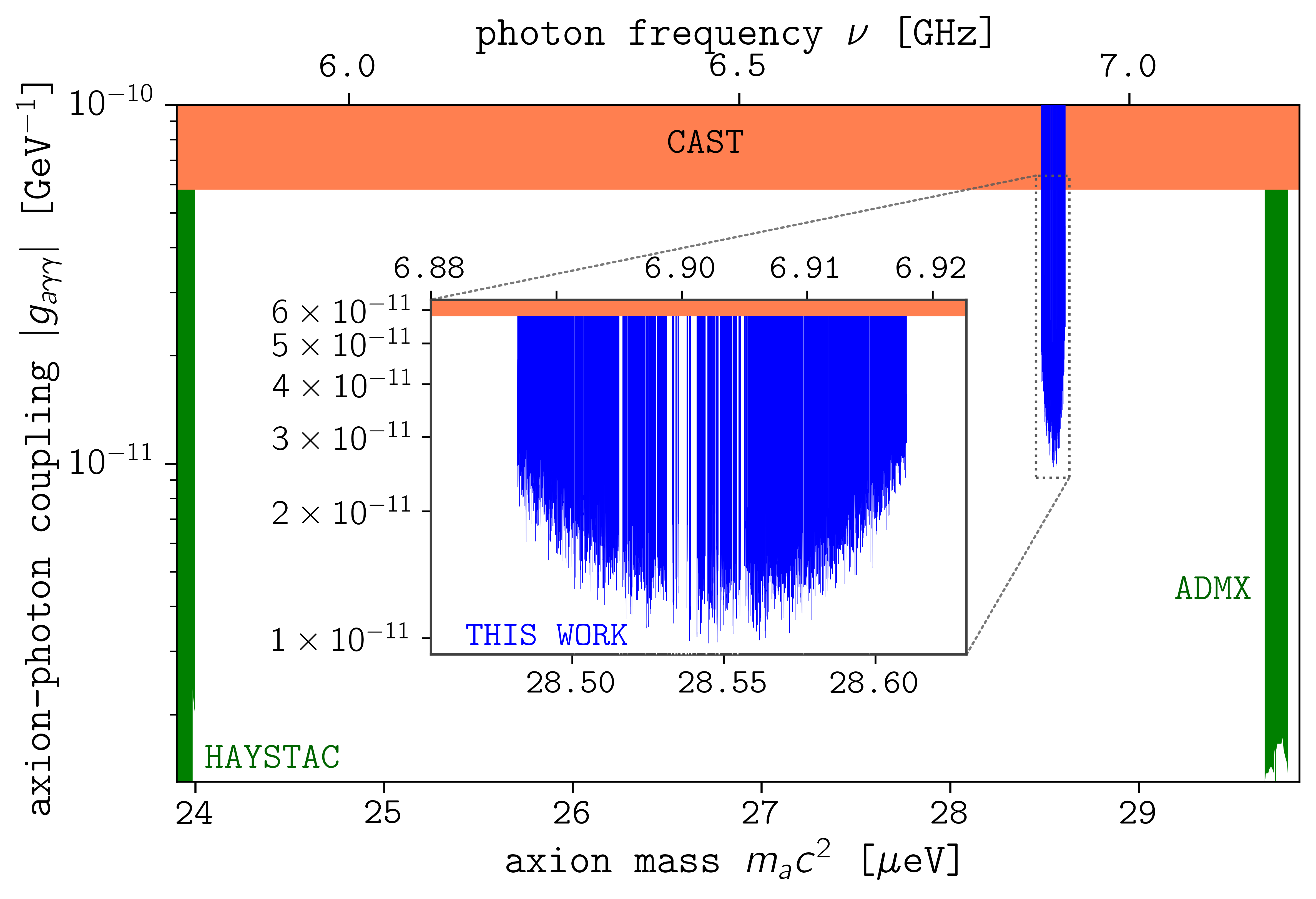}
    \caption{New exclusion region (blue) at 90\% CL. Narrow gaps indicate masked bins and vetoed neighborhoods around above-threshold excursions. Other laboratory-based constraints in the vicinity of the newly excluded region are also shown \cite{PhysRevLett.133.221005, PhysRevD.97.092001, PhysRevLett.118.061302, PhysRevLett.121.261302}.}
    \label{fig_2}
\end{figure}

The main results of this work are depicted in Fig.~\ref{fig_2}. We analyzed
12960 files of 10.0~s each (total integration time 36.0~h) in the
6.886909--6.917922~GHz band (28.4820--28.6102~$\upmu$eV), with a 
sensitivity peak of $9.718\times10^{-12}\,\mathrm{GeV}^{-1}$ at 6.902112 (28.5448~$\upmu$eV) sustained over a full $K_g$-bin resolution element.
We excluded axions at 90\% CL in the unmasked region
comprising 9555 of 10841 GS bins (88.1\%). The candidates identified for rescanning in Table \ref{TableII} will be re-evaluated during a future experimental run using upgraded hardware.

This proof-of-concept test establishes new exclusion limits, marks a first step toward broadband operation of the DALI(PoP) prototype, and lays the groundwork for the full-scale DALI experiment, designed to complement existing axion DM searches by probing a region of parameter space that, while well motivated by postinflationary scenarios, remains poorly explored because QCD-axion searches at high frequencies are technically challenging.

\section*{Acknowledgements} Thanks to J. Potticary and R. Barreto for technical support. The project that gave rise to these results received the support of a fellowship from “la Caixa” Foundation (ID 100010434). The fellowship code is LCF/BQ/PI24/12040023. J. D M. acknowledges support from the Spanish Ministry of Science, Innovation and Universities and the Agency (EUR2024-153552 financed by MICIU/AEI/10.13039/501100011033). We gratefully acknowledge financial support from the Severo Ochoa Program for Technological Projects and Major Surveys 2020-2023 under Grant No. CEX2019-000920-S; Recovery, Transformation and Resiliency Plan of Spanish Government under Grant No. C17.I02.CIENCIA.P5; Operational Program of the European Regional Development Fund (ERDF) under Grant No. EQC2019-006548-P; IAC Plan de Actuación 2022. J.F.H.-C. was supported by the Resident Astrophysicist Programme of the Instituto de Astrofísica de Canarias. This work was supported by the RIKEN's program for Special Postdoctoral Researchers (SPDR). This work is part of Grant No. CEX2025-001609-S, awarded to the Instituto de Astrofísica de Canarias under the Severo Ochoa Centre of Excellence program and funded by MICIU/AEI/10.13039/501100011033. M.M. acknowledges funding by the Deutsche Forschungsgemeinschaft (DFG, German Research Foundation) under Germany’s Excellence Strategy---EXC 2121 "Quantum Universe"---390833306. \newline

J. D M., E.J., E.H-S., J.F.H-C., H.L-H., and D.C. conceived, designed and built the prototype. J. D M., E.J., and D.C. installed and operated the detector. J. D M. and E.J. organized the data-taking runs and general operation of the experiment, coordinated and monitored the data taking. J. D M. supervised the project, processed and analyzed the data, and led the article drafting and editing with input from all authors. All authors provided critical feedback and helped shape the research, analysis, and manuscript.

\section*{Data availability} The data that support the findings of this article are not publicly available. The source data and code used to generate the plots in this paper are available from the corresponding author upon reasonable request.


\bibliography{apssamp}






\end{document}